\documentclass[aps,pra,twocolumn,showpacs,superscriptaddress,groupedaddress]{revtex4}
\usepackage{dsfont}
\usepackage{amsthm}
\usepackage{amsmath}
\usepackage{amssymb}
\usepackage{esint}
\usepackage{graphicx}
\usepackage{mathrsfs}             % See geometry.pdf to learn the layout options. There are lots.
\usepackage{graphicx}
\usepackage{amssymb}
\usepackage{epstopdf}
\usepackage{textpos}
\usepackage{color}
\usepackage{lipsum}
\usepackage{xcolor}
\newcommand{\ket}[1]{|#1 \rangle}

\newcommand{\mean}[1]{\langle #1 \rangle}

\newcommand{\tr}[1]{\mathrm{tr}\left\{#1\right\}}

\newcommand{\mc}[1]{\mathcal{#1}}

\begin{document}
\title{Quantum Virtual Cooling}
\date{\today}

\author{Jordan Cotler}\email{jcotler@stanford.edu}

\affiliation{\it Stanford Institute for Theoretical Physics, Stanford University, Stanford, CA 94305, US}

\author{Soonwon Choi}

\affiliation{\it Department of Physics,
University of California, Berkeley, CA 94720, USA}

\author{Alexander Lukin}

\affiliation{\it Department of Physics, Harvard University, Cambridge, MA 02138, USA}

\author{Hrant Gharibyan}

\affiliation{\it Stanford Institute for Theoretical Physics, Stanford University, Stanford, CA 94305, US}

\author{Tarun Grover}

\affiliation{\it Department of Physics, University of California at San Diego, La Jolla, CA 92093, USA}

\author{M. Eric Tai}
\affiliation{\it Department of Physics, Harvard University, Cambridge, MA 02138, USA}

\author{Matthew Rispoli}
\affiliation{\it Department of Physics, Harvard University, Cambridge, MA 02138, USA}

\author{Robert Schittko}
\affiliation{\it Department of Physics, Harvard University, Cambridge, MA 02138, USA}

\author{Philipp M. Preiss}
\affiliation{\it Department of Physics, Harvard University, Cambridge, MA 02138, USA}
\affiliation{\it Physics Institute, Heidelberg University, 69120 Heidelberg, Germany}

\author{Adam M. Kaufman}
\affiliation{\it Department of Physics, Harvard University, Cambridge, MA 02138, USA}

\affiliation{\it JILA, National Institute of Standards and Technology and University of Colorado,
and Department of Physics, University of Colorado, Boulder, Colorado 80309, USA}
\author{Markus Greiner}

\affiliation{\it Department of Physics, Harvard University, Cambridge, MA 02138, USA}

\author{Hannes Pichler}

\affiliation{\it ITAMP, Harvard-Smithsonian Center for Astrophysics, Cambridge, MA 02138, USA}

\affiliation{\it Department of Physics, Harvard University, Cambridge, MA 02138, USA}

\author{Patrick Hayden}

\affiliation{\it Stanford Institute for Theoretical Physics, Stanford University, Stanford, CA 94305, US}

%Authors, in no particular order:
%\author{Jordan Cotler}
%\affiliation{Stanford Institute for Theoretical Physics, Stanford University, Stanford, CA 94305}
%Soonwon Choi
%Alexander Lukin
%Hannes Pichler
%Markus Greiner
%Patrick Hayden
%Hrant Gharibyan
%Tarun Grover

\begin{abstract}
We propose a quantum information based scheme to reduce the temperature of quantum many-body systems, and access regimes beyond the current capability of conventional cooling techniques. We show that collective measurements on multiple copies of a system at finite temperature can simulate measurements of the same system at a lower temperature. This idea is illustrated for the example of ultracold atoms in optical lattices, where controlled tunnel coupling and quantum gas microscopy can be naturally combined to realize the required collective measurements to access a lower, virtual temperature. Our protocol is experimentally implemented for a Bose-Hubbard model on up to 12 sites, and we successfully extract expectation values of observables at half the temperature of the physical system.  Additionally, we present related techniques that enable the extraction of zero-temperature states directly. 

% Given two copies of a many-body system at temperature $T$, we propose protocols to measure correlation functions of a single copy at temperature $T/2$, half the temperature of the physical systems.  Our protocols leverage the quantum entanglement between the systems and their surroundings to simulate a virtual system at half of the physical temperature.  The details of the protocols do not depend on the physical temperature of the physical system, and so can be used to obtain additional (virtual) cooling after available physical cooling methods have been deployed.  We also present similar techniques for extracting the zero-temperature state more directly.  We detail implementations tailored to cold-atom systems on optical lattices, and demonstrate the viability of our methods with experimental data from a Bose-Hubbard model.  From the experimental data, we can successfully extract correlation functions at half the temperature of the physical systems.  Finally, we explore other applications, such as using our techniques to virtually probe finite-temperature phase transitions.

\end{abstract}


\maketitle

%\emph{Introduction} ---

\section{Introduction}

Quantum simulators have been proposed to understand the complex properties of strongly correlated quantum many-body systems \cite{Lloyd1996,Gross2017,Georgescu2014}. Significant progress has been made in building both analog and digital quantum simulators with a variety of quantum optical systems \cite{Jurcevic2014,Zhang2017,Guardo2018,Lienhard2018,Bernien2017,Barends2016,Eichler2015,Deverot2013}. A particularly successful approach is to use cold neutral atoms in optical lattices to emulate the physics of interacting electrons in solid state systems \cite{Jaksch1998, Greiner2002, Kohl2005, Aidelsburger2013, Miyake2013, Mancini2015,Stuhl2015,Baier2016, Gross2017}. This is exemplified by recent experimental advances that enable explorations of quantum magnetism \cite{Simon2011, Greif2013, Hart2015,Parsons2016, Boll2016, Cheuk2016, Mazurenko2017}, measurements of many-body entanglement \cite{Islam1, Kaufman2016, Brydges2018}, and studies of quantum dynamics out of equilibrium with bosonic and fermionic atoms \cite{Chenau2012, Meinert2014, MBLpaperBloch, Kaufman2016, ALukin1}. 

One of the central, outstanding challenges in these experiments is to reach the low temperatures needed to access strongly correlated phases. A prominent example is given by the doped Fermi-Hubbard model with cold atoms, where small energy scales lead to correspondingly stringent temperature requirements \cite{Gross2017}. Even though recent progress in reducing temperatures (e.g. via entropy redistribution techniques \cite{Mazurenko2017,Chu2018,Ho2009,Kantian2018}) allows current quantum simulators to compete with the most advanced quantum Monte Carlo algorithms on classical computers \cite{Gross2017}, the observation of extremely low temperature phenomena such as d-wave superconductivity remains elusive. This calls for the development of new techniques to reduce temperatures in quantum simulators. 
%Several methods to effectively cool such systems have been proposed and implemented, including entropy reLdistribution [CITE] and sub-waveletgh lattices [CITE]. 

\begin{figure}[t]
  \centering
  \includegraphics[width=0.48\textwidth]{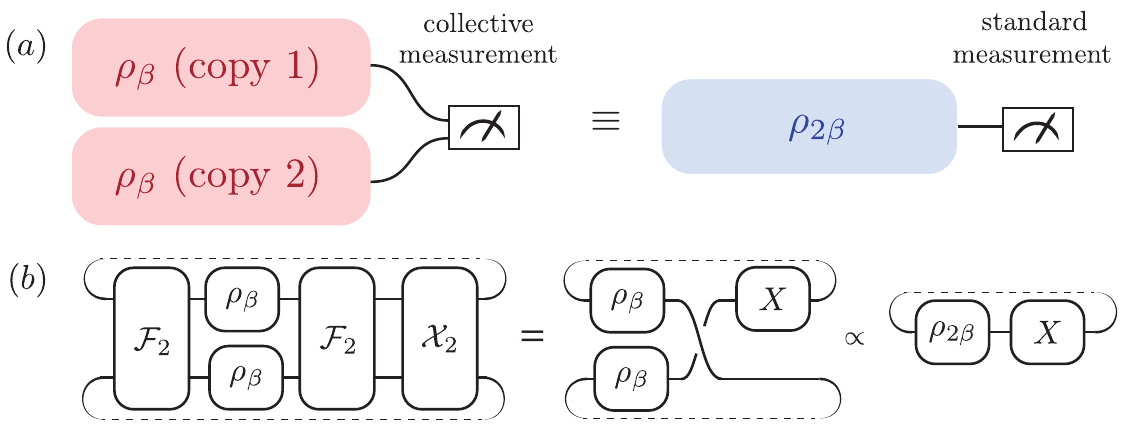}
  \caption{(a) Schematic representation of the virtual cooling protocol. Collective measurements on two copies of a thermal state $\rho_\beta$ at temperature $T=1/(k_B \beta)$ correspond to standard measurements at half the temperature, $T/2$.
  (b) Diagrammatic representation. Two copies are evolved with the unitary $\mc{F}_2$, and a subsequent measurement of $\mc{X}_2$ and $\mc{R}_2$ is performed. In combination this gives the expectation value $\text{tr}\{\rho_\beta \, \rho_\beta \, X\} \propto \text{tr}\{\rho_{2 \beta} X\}$ corresponding to half the original temperature.  We can also measure the proportionality constant with a similar procedure.}
% which is $1/\text{tr}\{\rho_{2\beta}\}$, by performing the same procedure with $X = \textbf{1}$.
  \label{fig1}
\end{figure}

In this work, we develop a novel approach to address this issue by introducing a measurement scheme that enables to access system properties at fractions of its actual temperature $T$. Importantly, our approach achieves this without the need to physically cool the system. Instead, our ``virtual'' cooling protocol to a temperature $T_{\rm virtual}=T/n$ ($n=2,3,\dots$) is facilitated by joint measurements on $n$ copies of the system at temperature $T$. For a schematic illustration see Fig.~\ref{fig1}(a). Our method can thus be used to virtually reduce the temperature of a system after all available physical cooling methods have been deployed.

Further, we detail implementations tailored to cold-atom systems in optical lattices, and illustrate our protocol in an experimental quantum simulation of the Bose-Hubbard model. Finally, we show how these ideas can be generalized and discuss protocols to distill the many-body ground state from multiple copies of thermal many-body states.

\section{Theoretical Overview}

In this section we present the theoretical ideas behind quantum virtual cooling, and discuss experimental implementation in the following sections.  We are interested in quantum many-body systems described by a thermal state $\rho(T)=e^{-\beta H}/Z$, where $H$ is the Hamiltonian of the system and $Z(T)=\text{tr}\{e^{-\beta H}\}$ is the partition function at inverse temperature $\beta=1/(k_BT)$. The measurement of an observable $X$ in the state $\rho$ gives the expectation value $\mean{X}_T= \tr{X\rho}$. Below we will discuss a protocol that allows us to effectively measure $\mean{X}_{T/n}$. The central idea is based on the ability to express the thermal density operator at $T/n$ by the $n$-th power of $\rho(T)$
\begin{equation}\label{eq1}
\rho(T/n)=\rho(T)^n/\text{tr}\{\rho(T)^n\}.
\end{equation}
In order to access the higher powers of the thermal state, we require $n$ copies of the state $\rho(T)$ prepared in parallel as well as the capability to implement operations that exchange the $n$ copies. 
More specifically, we have $\tr{X\rho^n}=\tr{X_s S_n \rho^{\otimes n}}$, where $S_n$ cyclically permutes quantum states in the $n$ copies, i.e. $S_n\ket{\psi_1}\otimes\ket{\psi_2}\otimes\dots\otimes\ket{\psi_n}=\ket{\psi_2}\otimes\ket{\psi_3}\otimes\dots\otimes\ket{\psi_1}$, and $X_s$ is the symmetrized embedding of $X$ on the $n$-fold replicated Hilbert space $X_s=\frac{1}{n}\sum_{m=1}^n S_n^m (X\otimes \textbf{1}^{\otimes (n-1)}) {S_n^m}^\dag$. %\footnote{Here $X$ is an arbitrary system operator. We embed it in the $n$-fold replicated Hilbert space as $X\otimes 1^{\otimes n-1}$.}. 
Therefore, the virtual measurement of $\mean{X}_{T/n}$ at temperature $T/n$ via Eqn.~\eqref{eq1} can be reduced to determining the expectation values $\mean{X_sS_n}$  and $\mean{S_n}$ on the $n$ copies of the state at temperature $T$.  This is illustrated in Fig. \ref{fig1}(b).  Measurements of expectation values of $S_n$ can be achieved with auxiliary qubits \cite{Ekert2002, Brun2004}, or directly via many-body state interferometry \cite{Alves2004,Daley1,Pichler1}, as recently demonstrated with cold atoms \cite{Islam1}. We also note that our protocols apply to subsystems which are thermal, even if the global system is not thermal.  In our experiments below, we leverage `eigenstate thermalization' \cite{deutsch1991,srednicki1994chaos, srednicki1998,rigol2008, rigol2016} to obtain thermal reduced density matrices from globally pure states of finite energy density in a chaotic system.  Earlier theoretical work provided numerical evidence that a chaotic eigenstate or a reduced density matrix of a thermal state encodes correlations at all temperatures \cite{Singh1, Grover1}.
% In fact, as demonstrated  in Ref. \cite{Grover1}, one can numerically obtain correlation functions of local operators at any temperature from a single pure state of a chaotic system by employing expressions analogous to Eqn.~\eqref{eq1}.

Below, we discuss protocols to measure $\mean{X_sS_n}$ for arbitrary $n$ and detail the procedure for the simplest example $n= 2$. We first focus on an interferometric measurement scheme and demonstrate that it can be implemented in current experiments with cold atoms. Alternative virtual cooling schemes using ancillary atoms are discussed below.  Finally, we show that schemes with ancillary atoms can be generalized to not only virtually cool a many-body system, but directly distill and prepare the many-body ground state from a thermal state. 
 Importantly, all of the discussed protocols are agnostic to the temperature $T$ of the physical system, and thus can be used to obtain additional, virtual cooling even after all available physical cooling methods have been deployed. 

% \textit{Interferometric measurement ---}

\section{Interferometric measurement}\label{IM}

To simplify the presentation we first discuss a virtual cooling scheme for bosonic atoms in optical lattices. 
%To simplify the presentation we first discuss a virtual cooling scheme that halves the original temperature of bosonic atoms.
The key idea is to represent the permutation operator $S_n$ in the bosonic Hilbert space as $S_n=\mc{F}_n^\dag \mc{R}_n\mc{F}_n$\,, where the unitary $\mc{F}_n$ denotes the discrete Fourier transformation 
\begin{align}
\label{Fourier1}
\mc{F}_n a_{p,j}\mc{F}_n^\dag=\frac{1}{\sqrt{n}}\sum_{k=1}^n e^{i\frac{2\pi kp}{n}}a_{k,j}
\end{align}
and $\mc{R}_n=\prod_j e^{-i2\pi/n\sum_{p=1}^n p\, n_{p,j}}$ \footnote{Note that a particle number superselection rule is required for Eqn.~\eqref{Fourier1}.}. Here $a_{p,j}$ denotes the bosonic annihilation operator on site $j$ in copy $p$, and $n_{p,j}=a_{p,j}^\dag a_{p,j}$ is the corresponding number operator. Note that $\mc{F}_n$ can be realized by simply introducing tunnel coupling between neighboring copies \cite{Daley1}, and $\mc{R}_n$ can be directly measured with a number-resolving quantum gas microscope. This representation of the permutation operator suggests that we introduce an operator $\mc{X}_n=\mc{F}_n X_s\mc{F}_n^\dag$, which is the discrete Fourier transform of the observable $X$ that we want to measure. With this definition we can express
\begin{align}\label{interferometer}
\mean{X}_{T/n}=\tr{\mc{X}_n\mc{R}_n (\mc{F}_n\rho^{\otimes n}\mc{F}_n^\dag)}/\tr{\mc{R}_n (\mc{F}_n\rho^{\otimes n}\mc{F}_n^\dag)}.
\end{align}
A measurement of $X$ at the virtually reduced temperature $T/n$ thus consists of a measurement of $\mc{X}_n\mc{R}_n$ and $\mc{R}_n$ after application of the discrete Fourier transform across the copies. For many interesting observables one finds $[\mc{X}_n,\mc{R}_n]= 0$ so that $\mc{R}_n$ and $\mc{X}_n$ can be measured independently.

As a specific example, we consider the experimentally simplest case $n=2$ and the measurement of the on-site density by choosing $X\equiv n_j$. The corresponding protocol consists of three steps. (i) We prepare $n=2$ identical
instances of the thermal many-body state $\rho(T)$. This can be achieved, for example, by preparing two identical states in neighboring 1D tubes, or 2D planes. It is essential that the copies are decoupled at this stage, which can be achieved by using a large optical potential between the tubes or planes to suppress any inter-copy tunneling.
 (ii) We then freeze the dynamics within each copy, and lower the potential between the two copies, e.g. using an optical superlattice. This induces tunneling between the two copies via the Hamiltonian $H_{\rm BS}=-J_{\rm BS}\sum_{j}(a_{1,j}^\dag a_{2,j}+\rm h.c.)$, which allows us to realize the so-called beamsplitter operation $\mc{F}_2$ that maps $\rho^{\otimes 2} \to \mathcal{F}_2 \rho^{\otimes 2} \mathcal{F}_2^\dag$.  Interactions between the atoms need to be turned off (e.g. via a Feshbach resonance) or made negligible as compared to $J_{BS}$ during this step.  (iii) Finally, we measure the on-site occupation number on all sites in both copies using a number-resolving quantum gas microscope. This gives direct access to $\mc{R}_2=(-1)^{\sum_{j}n_{1,j}}$ and $\mc{X}_2=\mc{F}_2 \frac{1}{2}(n_{1,j}+n_{2,j})\mc{F}_2^\dag=\frac{1}{2}(n_{1,j}+n_{2,j})$. Averaging the results over multiple experiments gives the expectation value of the local density at $T/2$ via Eqn.~\eqref{interferometer} (for a schematic of a single measurement trial, see Fig. \ref{fig2}(a) below).  Remarkably, this experimental procedure parallels the one employed to determine the second order R\'{e}nyi entropy of cold atoms, with atom number-resolved measurements being the only additional requirement.  Such measurements were first demonstrated for one-dimensional systems using full-atom-number-resolved imaging in quantum gas microscope \cite{Kaufman2016}.

% As an aside, we note that expressions of the form in Eqn.~\eqref{interferometer} can be measured using an alternative, more direct method in the special case that $\rho$ is a reduced density matrix obtained from a pure state that can be expanded with positive coefficients in the measurement basis, i.e., $\rho(A, A') = \sum_{B} \psi(A, B) \psi(A', B)$ where $\{A\}, \{B\}$ denote the complete basis for regions $A,B$ while $\psi$ is real and positive. As an example, when $n = 2$,  $\mean{X}_{T/2}$ is given by 
% \begin{equation}
%\frac{\sum_{A,B, A',B'} \psi(A,B) \psi(A',B') \psi(A',B) \psi(A,B') X(A,A)}{\sum_{A,B, A',B'} \psi(A,B) \psi(A',B') \psi(A',B) \psi(A,B') }
% \end{equation}
%where for simplicity we have assumed that the operator $X$ is diagonal in the measurement basis. The above expression can be evaluated via Monte Carlo methods by sampling $X(A,A)$ with the probability distribution function $\psi(A,B) \psi(A',B') \psi(A',B) \psi(A,B')$ defined on the two copies of the system. Note that, unlike the interferometric measurement discussed above, here one does not evaluate the numerator and the denominator separately, and therefore, the measurement scales more favorably with increasing system size. Physically, systems which fall under this special class do not suffer from `sign problem', e.g., the ground state of boson-Hubbard model with unfrustrated hopping.
 
\section{Experimental demonstration}

\begin{figure}[t]
	\centering
	\includegraphics[width=0.48\textwidth]{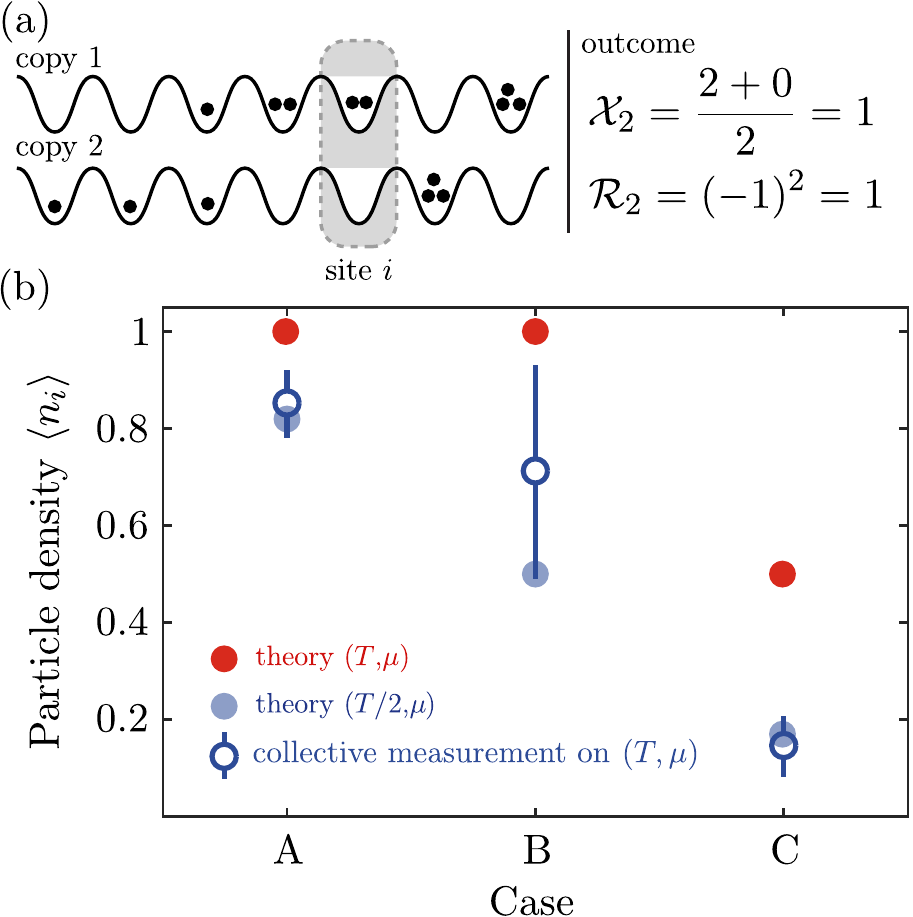}
	\caption{(a) Schematic for a single measurement trial of $\mathcal{X}_2 = \frac{1}{2}(n_{1,i} + n_{2,i})$ and $\mathcal{R}_2$ restricted to the $i$th site, after $\mathcal{F}_2$ has been applied to the two copies. (b) Measured single-site density, averaged over all but the edge sites of the chain, after virtual cooling has been applied to the system (blue circles with vertical error bars). Red discs show the single-site density of the state before our protocols are utilized (the actual density of particles in each experiment), whereas light blue discs correspond to the prediction of the effective thermal ensemble (see text) at half the temperature. Agreement of the data with the reduced-temperature ensemble validates the applicability of our method in the experimental system.  Error bars denote the standard error of the mean.}
	\label{fig2}
\end{figure}
In order to demonstrate our protocol, we experimentally realize it in a one-dimensional Bose-Hubbard model. 
%an existing experimental data that have been previously obtained in Ref.~\cite{Kaufman2016}.
In the experiment, a Bose-Einstein condensate of $^{87}$Rb atoms is loaded into a two-dimensional optical lattice positioned at the focus of a high-resolution imaging system. The dynamics of the atoms is well-described by a Bose-Hubbard Hamiltonian parametrized by tunneling strength $J$ and on-site interaction energy $U$ (see Ref.~\cite{Kaufman2016} for details).

The experimental protocol consists of four steps: initialization, quenched thermalization dynamics, beamsplitter operations, and measurements. During initialization, optical potentials are sequentially manipulated in order to isolate an initial product state, $\ket{\psi_0}$, with a single atom on the central $2\times 6$ sites of a $2\times L$ plaquette in the deep $45 \mathrm{E_r}$ lattice where the tunneling between the sites is negligible~\cite{Kaufman2016}. Each $1\times L$ tube represents an identical copy of the system. Next, the lattice potential along the chains is suddenly lowered, allowing particles to tunnel and interact within each chain. It has been previously shown \cite{Kaufman2016} that this quenched dynamics drives the thermalization of small subsystems within the chain. Hence, after sufficiently long time evolution, the state of the subsystem can be described by an effective temperature $T$ and chemical potential $\mu$, which are determined by the total energy and particle number density of $\ket{\psi_0}$. (See also \cite{Grover1}.)  After the desired time evolution the dynamics of the system is frozen by suddenly increasing the lattice depth along the chains, and a beamsplitter operation $\mathcal{F}_2$ is implemented by lowering the potential barrier between the two chains, such that particles can tunnel (in the transverse direction) for a prescribed time. Finally, the number of particles on each individual lattice site is measured. This procedure is repeated multiple times in order to obtain sufficient statistics.

We apply our virtual cooling protocol in three regimes (A, B, and C), with differing initial states $\ket{\psi_0}$, system size $L$, and Hamiltonian parameters $U/J$. For the data sets A and B, each of $L=6$ sites is initially occupied by one particle, whereas for the data set C, only the middle six out of the total $L=12$ sites are occupied by one particle per site.
The tunneling rates are set such that $U/J \approx 1.56$ (data set A) or $0.33$ (data sets B and C).
These combinations lead to the effective temperatures and chemical potentials $(T/J,\mu/J) \approx (3.5,-1.0)$, $(11.5,-6.3)$, and $(18.3,-17.7)$ of subsystems for data sets A, B, and C, respectively.
%(see Methods) \textcolor{red}{[JC: Should I comment this out for now?]}.
Based on our protocol, we extract the average particle number density $\langle n_i \rangle$ of the $i$th site for thermal ensembles at reduced temperature. 
 
Fig.~\ref{fig2}(b) shows the resulting single-site particle density after virtual cooling for all three cases. We compare these results with the initial single-site density at the original temperatures as well as theoretical predictions from an ideal thermal ensemble $\rho_{2\beta}$ at half of the original temperatures.  All data points are in good agreement with the reduced temperature ensemble indicating that our virtual cooling scheme works in the experimental system.

\section{Observables}
The protocol presented in Sec.~\ref{IM} allowed us to measure local densities at reduced temperatures. In this section we discuss some of the issues that arise when generalizing this scheme to more complicated observables, and present an alternative protocol that avoids these issues. 

One of the useful properties of the single-site density $X \equiv n_j$ is that its symmetrized version $X_s = \frac{1}{2}(n_{1,j}+n_{2,j})$ is invariant under conjugation by $\mc{F}_2$, i.e, $\mc{X}_2=\mc{F}_2 X_s\mc{F}_2^\dag=\frac{1}{2}(n_{1,j}+n_{2,j})$. Thus, $\mathcal{X}_2$ is easily measured by averaging the number of atoms on the $j$th site in the two copies.  Furthermore, $\mathcal{X}_2$ commutes with $\mathcal{R}_2$, and so we can measure the observables in either order.  In fact, $\mathcal{X}_2$ and $\mathcal{R}_2$ commute with all single-site densities $n_{1,j}$, $n_{2,k}$, and so we can simply measure the individual particle numbers and combine them to compute the expectation values of $\mathcal{X}_2$ and $\mathcal{R}_2$.

For more complicated observables such as density-density correlators $X\equiv n_jn_\ell$, the situation is more subtle. A direct application of the procedure outlined above requires a measurement of
\begin{align}
\label{nn1}
\mc{X}_2 &= \mc{F}_2 \frac{1}{2}(n_{1,j}n_{1,\ell}+n_{2,j}n_{2,\ell})\mc{F}_2^\dag \nonumber \\
&=\frac{1}{4}\left({n}_{1,j} + {n}_{2,j} \right)\left({n}_{1,\ell} + {n}_{2,\ell} \right) \\
& \qquad + \frac{1}{4}\left(a_{1,j}^\dagger a_{2,j} + a_{2,j}^\dagger a_{1,j} \right)\left(a_{1,\ell}^\dagger a_{2,\ell} + a_{2,\ell}^\dagger a_{1,\ell} \right)\,. \nonumber
\end{align}  
While the first term in Eqn.~\eqref{nn1} (i.e., the final equality) is easily measurable with standard quantum gas microscopy, the second term requires additional interferometric apparatus.

Before proceeding with the discussion of an alternative protocol that avoids this issue (see Sec.~\ref{Ancilla}), let us note that the first term of Eqn.~\eqref{nn1} by itself contains interesting information about the system at half of its temperature.  This first term of Eqn.~\eqref{nn1} is easy to measure, since it commutes with $\mathcal{R}_2$ and all of the number operators.  Doing so would output the unconventional correlator
\begin{equation}
\label{unconventional1}
\frac{1}{2} \, \text{tr}\{n_j n_\ell \, \rho(T/2)\} + \frac{1}{2} \, \frac{\text{tr}\{n_j \, \rho(T) \, n_\ell \, \rho(T)\}}{\text{tr}\{\rho(T)^2\}}\,.
\end{equation}
The term on the left here is the desired equal-time density-density correlator at half the system temperature, whereas the term on the right is peculiar.  In fact, this peculiar term is equal to the \textit{unequal imaginary-time} correlator $\frac{1}{2} \, \text{tr} \{n_j(1/T) \, n_\ell \, \rho(T/2)\}$ where $n_j(\tau) = e^{H \tau} n_j e^{-H \tau}$ is the number density evolved in imaginary time.  If our system is translation invariant and at sufficiently low temperature, we expect $\text{tr}\{n_j n_\ell \, \rho(T/2)\}$ to depend on $|j-\ell|$, whereas the peculiar term should not strongly depend on $|j-\ell|$. This is because at low temperatures, the large imaginary time evolution of the operator $n_j$ scrambles it strongly, destroying the memory of its initial position $j$.  Indeed, in the limit of $T \to 0$, the peculiar term is just $\langle \psi_0| n_j |\psi_0\rangle \langle \psi_0|n_\ell |\psi_0\rangle$ which is clearly independent of $|j - \ell|$.  At high temperature and small $|j-\ell|$, both terms in Eqn.~\eqref{unconventional1} have a nontrivial dependence on $|j - \ell|$ and so we are unable to extract each term separately. Nevertheless, it is interesting to note that our protocol yields some information about the unequal imaginary-time correlator in this regime.  We note also that when $|j-\ell|$ is much larger than the thermal correlation length, both terms in Eqn.~\eqref{unconventional1} approach $\frac{1}{2} \, \text{tr}\{n_j \, \rho(T/2)\} \, \text{tr}\{n_\ell \, \rho(T/2)\}$.  We explore the dependence of $\text{tr}\{n_j \, \rho(T) \, n_\ell \, \rho(T)\}$ on $|j - \ell|$ as a function of $T$ in the Supplementary Materials, and confirm that there is essentially no dependence at sufficiently low temperatures.  

%Another approach to measuring density-density correlators is to use a different protocol involving an ancillary qubit to implement a controlled swap operation, which we will discuss below.  In this protocol, one does not need to measure $\mathcal{R}_2$ by observing the particle number distribution with a quantum gas microscope, and so we have more flexibility in our ability to measure $\mathcal{X}_2$.  Another interesting observable which is easy to measure in the ancillary qubit setting is the hopping operator $X \equiv a_j a_\ell^\dag + \text{h.c.}$, and accordingly $X_s = \frac{1}{2}(a_{1,j} a_{1,\ell}^\dag + a_{2,j} a_{2,\ell}^\dag) + \text{h.c.}$ which satisfies $\mathcal{X}_2 = X_s$.

\subsection{Ancilla Qubit Approach to Cooling}\label{Ancilla}

The example above shows that for some observables a direct measurement of the $\mathcal{F}_2$ conjugation may be challenging. Here we present an alternative approach which is experimentally feasible.

Consider a non-destructive measurement of the swap operator, $S_2$, on two systems which are each prepared in the state $\rho$. Since $S_2$ is unitary and hermitian, the two possible measurement outcomes are $\pm1$, corresponding to projections into the symmetric or anti-symmetric subspace with respect to the exchange of the two copies. The state after such a measurement is thus given by $\mc{P}_{\pm}(\rho{\otimes}\rho)\mc{P}_{\pm}/\tr{\mc{P}_{\pm}(\rho{\otimes}\rho)}$, with $\mc{P}_\pm=(1\pm S_2)/2$.
If the measurement outcome is $-1$, we discard both systems. But for those instances that yield a measurement $+1$ we retain one of the systems, and discard only the other one.  The resulting state of this first system $\rho_1$ is obtained by tracing out the degrees of freedom of the second system,
\begin{align}
\rho_1=\frac{{\rm{tr}}_{2}\{\mc{P}_{+}(\rho{\otimes}\rho)\mc{P}_{+}\}}{\tr{\mc{P}_{+}(\rho{\otimes}\rho)}}=\frac{\rho+\rho^2}{1+\tr{\rho^2}}\,.
\end{align}
For an initial thermal state $\rho(T)$, the new state $\rho_1$ corresponds to a mixture of $\rho(T)$ and $\rho(T/2)$.  The success probability for achieving $\rho_1$ is $p_+=(1+\tr{\rho^2})/2$ which is always larger than $1/2$.

Now to measure $\text{tr}\{X\,\rho(T/2) \}$, we first measure $\text{tr}\{X\, \rho(T) \}$ and then $\text{tr}\{X\,\rho_1 \}$.  Through the process of measuring $X$ with respect to $\rho_1$, we automatically determine $p_+$.  Then we put together our measurements as
\begin{equation}
\frac{2 p_+}{2 p_+ - 1}\,\text{tr}\{X \, \rho_1 \}  - \frac{1}{2 p_+ - 1} \, \text{tr}\{X \, \rho(T)\}= \text{tr}\{X\,\rho(T/2)\}
\end{equation}
which gives us the desired measurement of $\rho(T/2)$.

\begin{figure}[t]
  \centering
  % \label{fig:distillation1}
  \includegraphics[width=0.45\textwidth]{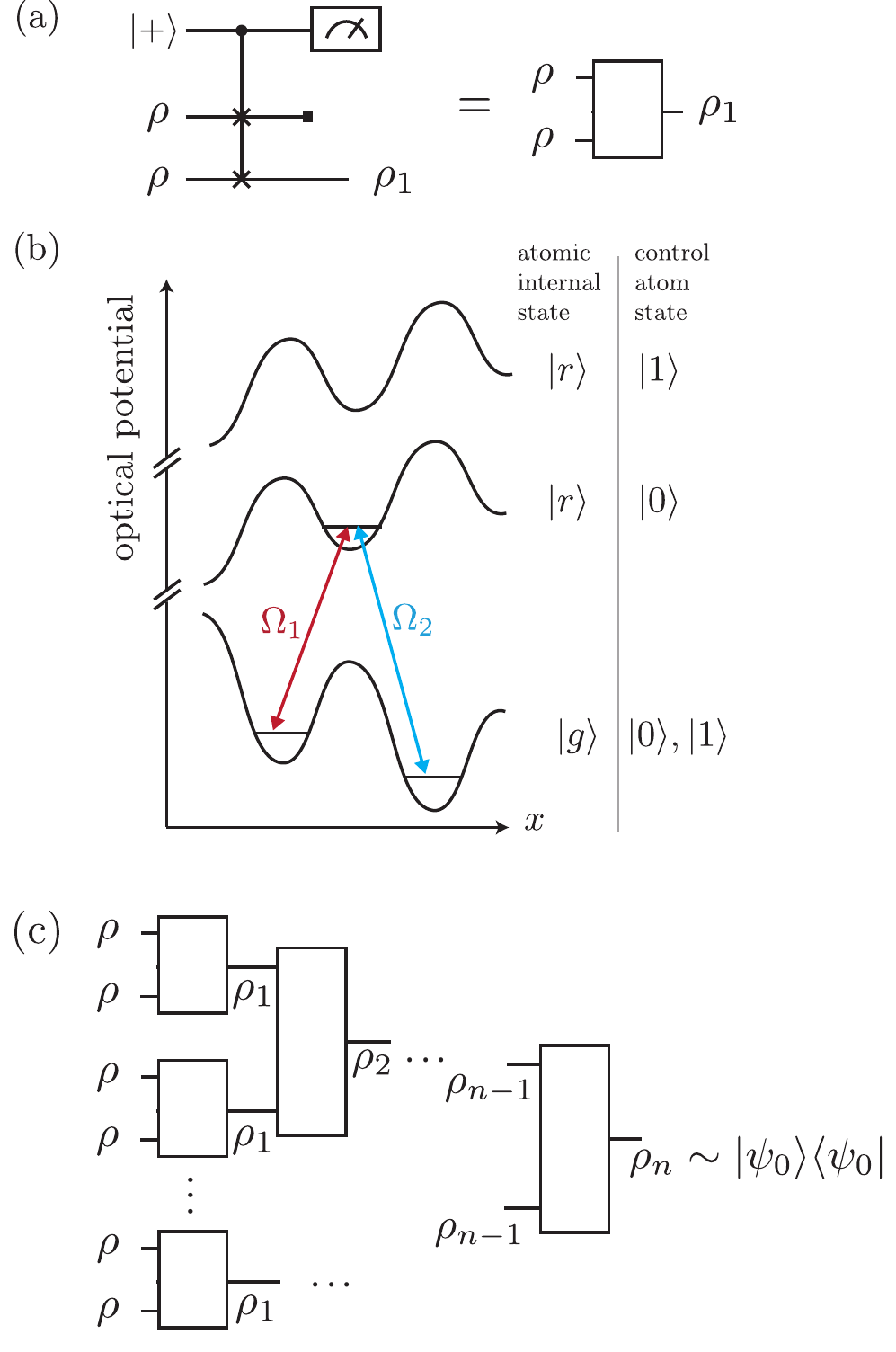}
  \caption{(a) Quantum circuit representation of ancilla qubit virtual cooling protocol.  Following the controlled swap of two copies of a quantum state, the control ancilla qubit is measured.  If the qubit is measured to be $|0\rangle$, then one of the system copies is discarded and the remaining system copy is in the state $\rho_1$.  (b) The controlled swap operation can be implemented for ultra-cold atoms on an optical lattice by the combination of photon-assisted hopping and the Rydberg blockade mechanism; excitation of the control atom in a Rydberg state conditionally prevents photon-assisted hopping. (c) Quantum circuit representation of ground state distillation protocol.  The protocol in (a) can be parallelized and nested, as shown in the diagram. If $\rho$ is a thermal state, then the ground state will be distilled.}
  \label{fig3}
\end{figure}

Non-destructive measurements of the swap operator are typically challenging. One way to realize such measurements is to use ancillary qubits \cite{Ekert2002}. A non-destructive measurements of the swap operator can then be realized by a simple quantum circuit, in which the ancillary qubit is initially prepared in the state $\tfrac{1}{\sqrt{2}}(\ket{0}+\ket{1})$, followed by the application of a controlled swap operation, exchanging the quantum state of the two copies conditional on the ancillary qubit being in state $\ket{0}$, and a final measurement of the ancillary qubit (see Fig.~\ref{fig3}).  (This is the opposite of the usual convention for controlled swap gates, but will be convenient immediately below.)

In a cold atom setup, one can envision realizing the required controlled swap operations using Rydberg interactions \cite{Pichler2}. For example, one might encode the ancillary qubit states in two internal states of an ancillary atom, with $\ket{0}$ being the internal ground state, and $\ket{1}$ a highly excited, long lived, Rydberg state. The Rydberg blockade mechanism can be used to control the tunnel coupling between two copies of an optical lattice and so realize a controlled exchange operation. To see this, consider inducing a tunnel coupling between the initially decoupled copies using a two photon Raman process  \cite{Jaksch2003,
Aidelsburger2013,
Miyake2013}. If this Raman process involves a Rydberg state as an intermediate level (see Fig.~\ref{fig3}), it is affected by the state of the ancillary atom. In particular, if the ancillary atom is in a Rydberg state, the strong dipolar interactions can lead to a shift of the energy of this intermediate state and completely inhibit tunneling. As a consequence, the exchange of the quantum state of the two copies can be completely controlled by the ancillary atom. We note that similar protocols have been discussed and analyzed in the literature \cite{Pichler2}.

 \section{Ground state distillation}
 
The ancilla qubit approach can be generalized to schemes that not only allow us to measure a system at reduced temperatures, but further enable the distillation of the ground state from multiple copies of a thermal ensemble. This is akin to entanglement purification proposals for quantum communication over noisy channels \cite{Duan1}. 

Consider again the protocol explained in the previous section, in which we couple $\rho \otimes \rho$ to an ancilla and obtain the state $\rho_1 = (\rho + \rho^2)/(1 + \text{tr}\{\rho^2\})$ with probability $p_+=(1+\tr{\rho^2})/2$.  If $\rho(T)$ is a thermal state at temperature $T$, then $\rho_1$ is a linear combination $\rho(T)$ and $\rho(T/2)$.  Clearly, $\rho_1$ has the same eigenvectors as $\rho$, but with different eigenvalues. In particular, $\rho_1$ is purer than $\rho$, and the eigenvalue of the largest eigenvector (i.e.~the ground state for thermal $\rho$) is larger. This purification is of course probabilistic, as its success is conditioned on the proper measurement outcome for $S_2$. Remarkably, the success probability $p_+=(1+\text{tr}\{\rho^2\})/2$ is always larger than $1/2$ and approaches $1$ as the system is purified. Starting with multiple copies one can iterate the above process, which will ultimately converge to a system in the largest eigenstate of $\rho$. For thermal states, the procedure distills the many-body ground state, i.e. the zero-temperature state.
%as depicted in Fig.~\ref{}.This process clearly converges to produce a system in the largest eigenstate of $\rho$. For thermal states this thus distills the many-body ground state.

\section{Limitations and Scalability}

We are often interested in local observables $X$, which in turn correspond to the local observables $X_s$.  Suppose that $X$ is supported on a subregion $R$.  Then $X_s$ is supported on the joint region $R_1 \cup R_2$ of the two corresponding system copies.
For concreteness, suppose our system is one-dimensional.
We desire to measure
$$\text{tr}\{X \, \rho(T/2)\} = \text{tr}_R\{X \, \rho_R(T/2)\}\,,$$
where $\rho_R(T/2) = \text{tr}_{\overline{R}}\{\rho(T/2)\}$ is the reduced density matrix of $\rho(T/2)$ on $R$.  Na\"{i}vely, it seems that we only need to perform our procedure on the subsystem $R_1 \cup R_2$ of the two copies.  However, this is not correct, since
$$\frac{\rho_R^2}{\text{tr}_R\{\rho_R^2\}} \not = \text{tr}_{\overline{R}}\left\{\frac{\rho^2}{\text{tr}\{\rho^2\}}\right\} = \text{tr}_{\overline{R}}\{\rho(T/2)\}\,.$$
Nonetheless, suppose we extend $R$ by buffering each of its boundaries by a number of sites corresponding to the correlation length of the system at temperature $T/2$.  Let us denote this extended region by $B$.  Here, $R \subset B$, but $B$ is smaller than the whole system.  The corresponding joint region of the two system copies is $B_1 \cup B_2$.  If we perform our procedure on the subsystem $B_1 \cup B_2$ of the two copies, we can access the state $\sigma_B \equiv \rho_B^2/\text{tr}_B\{\rho_B^2\}$, which satisfies $\sigma_B \approx \text{tr}_{\overline{R}}\{\rho(T/2)\}$, and therefore
$$\text{tr}_B\{X \sigma_B\} \approx \text{tr}\{X \, \rho(T/2)\}\,.$$
So if we choose $B$ large enough (but in most cases, smaller than the size of the entire system), we can still approximately measure our desired observable.

\subsection{Scalability}

We discuss the scalability in terms of two parameters, the temperature of the total system and the size of the subsystem to be measured. In particular, we are interested in the limit where the temperature is low and the total system size is large.  In an experiment, the performance of a measurement protocol is fundamentally limited by the number of repetitions required to determine the averages  achieve sufficiently high precision.  In our setting, the measurement statistics required to precisely measure the denominator $Z_n=\textrm{tr}\{\rho(T)^n\}$ in Eqn.~\eqref{eq1} may be a limiting factor. In a many-body system, $Z_n$ is directly related to the R\'enyi-$n$ entropy $S_n = \frac{1}{1-n} \log(Z_n)$, which scales with volume for local systems. $Z_n$ is therefore often exponentially small in the system size.

Hence, one would generally need a large number of measurements $N_m \sim 1/Z_n^2\sim \exp\{2\,s(T)|R|\}$, where $s(T)$ is the entropy density at temperature $T$ and $|R|$ is the size of the subregion on which $\rho(T)$ is supported. In the limit of low temperature, this scaling becomes favorable since $s(T)$ generally decreases. However, the thermal correlation length $\xi(T/n)$ can increase as $T$ is lowered, requiring a larger subregion size $|R|\geq \xi(T/n)$. Together, the number of measurements required to achieve some fixed precision scales as $N_m \sim \exp\{2\,s(T)\,\xi(T/n)\}$.  In practice, the correlation length of particular two-point functions may be smaller than the thermal correlation length, depending on the choice of operator insertions.  Accordingly, a smaller, effective correlation length for particular observables yields a more favorable scaling in the number of measurements.

Of course, if $\rho$ is only approximately thermal, then expectation values of $\rho^n/\text{tr}\{\rho^n\}$ for larger values of $n$ can have amplified deviations from thermality.  However, if we are interested in the physics of the ground state $|\psi_{0}\rangle$, then $\rho^n/\text{tr}\{\rho^n\} \sim |\psi_{0}\rangle \langle \psi_{0}|$ for larger values of $n$ so long as $|\psi_{0}\rangle$ is the dominant eigenstate of $\rho$.

\section{Discussion}

Reaching low temperatures is paramount for studying interesting quantum many-body phases with quantum simulators. In particular, the small energy scales in cold atom systems pose a major challenge for accessing the required temperature regimes. In this work, we proposed and demonstrated novel techniques that enable access to properties of a system at a fraction of its actual temperature. This virtual cooling is enabled by collective measurements on multiple copies of the system. 

More generally, our schemes illustrate a connection between thermal physics and entanglement. In particular, the temperature of a system is intimately connected to its entanglement with its surroundings \cite{srednicki1994chaos, srednicki1998, rigol2012, Grover1, deutsch2013,Kaufman2016}. Accordingly, measuring correlations of a thermal system at virtually lower temperatures involves manipulating and probing entanglement.  This is why the tools for measuring a system at virtually lower temperatures resemble those that allow access to entanglement entropies \cite{Alves2004,Daley1,Islam1}.

A natural future direction is to experimentally perform quantum virtual cooling for more complicated observables.  A particularly interesting application would be to experimentally study a quantum many-body system with a finite-temperature phase transition at some temperature $T_c$.  One could prepare the system at some temperature $T > T_c$, and use virtual cooling to probe features at or below the phase transition.  (For related theoretical work, see \cite{Fratus1}.)  Understanding the range of applicability of quantum virtual cooling is an exciting theoretical and experimental program, which will require new insights in subsystem ETH and thermalization.

%(see Ref.\cite{melko2011} for a quantum Monte Carlo demonstration of such an application).\\

% There are several interesting future directions suggested by our work.  In quantum annealing, one encodes the solution to the problem as the ground state of a Hamiltonian.  A state at finite temperature is prepared, and the temperature is (slowly) reduced until...

%\begin{itemize}
%\item Our new understanding can be leveraged experimentally
%\item Edge effects
%\item Future experiments
%\end{itemize}
\vskip.3cm
\noindent \textbf{Acknowledgments.\quad} We thank Alex Avdoshkin for helpful conversations.  JC is supported by the Fannie and John Hertz Foundation and the Stanford Graduate Fellowship program. SC acknowledges support from the Miller Institute for Basic Research in Science.  AL, RS and MG are supported by the NSF, the Gordon and Betty Moore Foundations EPiQS Initiative, and the Air Force Office of Scientific Research MURI program.  HG was supported in part by NSF grant PHY-1720397. MR was supported by an NSF Graduate Research Fellowship.  PMP acknowledges funding through the ERC consolidator grant 725636 and the Daimler and Benz foundation.  TG is supported as an Alfred P. Sloan Research Fellow.  HP is supported by the NSF through a grant for the Institute for Theoretical Atomic, Molecular, and Optical Physics at Harvard University and the Smithsonian Astrophysical Observatory. PH was supported by AFOSR (FA9550-16-1-0082), CIFAR and the Simons Foundation.
%\textcolor{red}{[Other acknowledgments here]}

\cleardoublepage
\onecolumngrid
\section*{Supplementary Materials}
\appendix
\section*{I. \quad Further Details of Quantum Virtual Cooling}
Here we present detailed quantum virtual cooling schemes, including ones that do not appear in the main text.  We analyze the case of two system copies, so that quantum virtual cooling allows us to probe observables at half of the physical temperature.  In particular, we specialize to bosons and fermions in optical lattices.

\subsection{Boson interferometry}

If our two identical systems are bosonic, then we can perform quantum virtual cooling along the lines of the main text.  In particular, we do not need an ancilla qubit to facilitate the application of the swap operator.  Consider the bosonic Hilbert space $\text{Sym}(\mathcal{H}_1 \otimes \mathcal{H}_2)$, comprising of two systems with $N$ sites each.  A basis for $\text{Sym}(\mathcal{H}_1 \otimes \mathcal{H}_2)$ is
\begin{equation}
|\{p_i\},\{q_j\}\rangle = \prod_{i,j=1}^N (a_{2,i}^\dagger - a_{1,i}^\dagger)^{p_i} (a_{2,j}^\dagger + a_{1,j}^\dagger)^{q_j}|\text{vac}\rangle
\end{equation}
for $\{p_i\}, \{q_j\} \in \mathbb{Z}_{\geq 0}^{\times N}$.  From Eqn.~\eqref{Fourier1}, $\mathcal{F}_{2}$ is a unitary which maps
\begin{align}
\mathcal{F}_{2} \,\frac{1}{\sqrt{2}}(a_{2,i}^\dagger + a_{1,i}^\dagger) \, \mathcal{F}_{2}^\dagger &= a_{2,i}^\dagger \\
\mathcal{F}_{2} \,\frac{1}{\sqrt{2}}(a_{2,i}^\dagger - a_{1,i}^\dagger) \, \mathcal{F}_{2}^\dagger &= a_{1,i}^\dagger\,.
\end{align}
% and let $\mathcal{F}_2 = \bigotimes_{i=1}^N \mathcal{F}_{2,i}$.  %As explained in the main text, the operator $\mathcal{F}_2$ can be implemented by making the two identical bosonic systems spatially adjacent to one another but separated by a barrier, and then allowing them to tunnel through the barrier.
Furthermore, $\mathcal{R}_2 = (-1)^{\sum_j n_{1,j}}$ is the total parity operator for the first of the two identical systems.  It is easy to check that
\begin{align}
S_2|\{p_i\},\{q_j\}\rangle &= \mathcal{F}_2^\dagger \mathcal{R}_2 \mathcal{F}_2 |\{p_i\},\{q_j\}\rangle\,,
\end{align}
and so
\begin{equation}
\text{tr}\{\mathcal{R}_2 \, \mathcal{F}_2 \, \rho^{\otimes 2} \, \mathcal{F}_2^\dagger \} = \text{tr}\{S_2 \, \rho^{\otimes 2}\} = \text{tr}\{\rho^2\}\,.
\end{equation}
Then if we have an operator $X$ that we wish to measure, the idea is to instead measure $\mathcal{X}_2 = \mathcal{F}_2 \frac{1}{2}(X \otimes \textbf{1} + \textbf{1} \otimes X) \mathcal{F}_2^\dag$ so that, in essence,
\begin{align}
\label{R2appEq1}
&\text{tr}\{\mathcal{R}_2 \, \mathcal{X}_2\,\mathcal{F}_2\, \rho^{\otimes 2}\, \mathcal{F}_2^\dagger \} = \frac{1}{2}\,\text{tr}\{\mathcal{R}_2 \,\mathcal{F}_2(X \otimes \textbf{1} + \textbf{1}\otimes X)\,\rho^{\otimes 2}\, \mathcal{F}_2^\dagger \} = \frac{1}{2}\,\text{tr}\{S_2 \, (X \otimes \textbf{1} + \textbf{1} \otimes X)\, \rho^{\otimes 2} \} = \text{tr}\{ X\,\rho^2\}\,.
\end{align}
Of course, there is a detailed measurement procedure which realizes the above equations.

To measure $\text{tr}\{X \, \rho^2\}/\text{tr}\{\rho^2\}$, we use the following procedure:
\begin{enumerate}
\item Start with the initial state $\rho^{\otimes 2}$.
\item Apply $\mathcal{F}_2$ to obtain
\begin{equation}
\sum_i \mathcal{F}_2 \, \rho^{\otimes 2} \, \mathcal{F}_2^\dagger.
\end{equation}
\item Measure the operator $\mathcal{X}_2$, given by
\begin{equation}
\label{form1}
\mathcal{X}_2 = \mathcal{F}_2\left(\frac{1}{2}\,X(\{a_{1,i}, a_{1,i}^\dagger\}) + \frac{1}{2}\,X(\{a_{2,i}, a_{2,i}^\dagger\})\right) \mathcal{F}_2^\dagger \,.
\end{equation}
Here, $X(\{a_{1,i}, a_{1,i}^\dagger\})$ denotes that the operator is written in terms of sums of products of creation and annihilation operators in the set $\{a_{1,i}, a_{1,i}^\dagger\}_{i \in \text{sites}}$, and similarly for $X(\{a_{2,i}, a_{2,i}^\dagger\})$.  The operator $\mathcal{X}_2$ has the property $[\mathcal{X}_2, \mathcal{R}_2] = 0$, which will be utilized shortly.  Suppose $\mathcal{X}_2 = \sum_i \lambda_i\,  P_i$ where the $\{P_i\}$ are orthogonal projectors.  Then after measurement one is left with
\begin{equation}
\sum_i P_i \,\mathcal{F}_2 \,\rho^{\otimes 2} \,\mathcal{F}_2^\dagger \, P_i\,.
\end{equation}
\item Measure $\mathcal{R}_2 = \Pi_+ - \Pi_-$ (where $\Pi_{\pm}$ is the projector onto the $\pm$ eigenspace) to obtain
\begin{equation}
\label{probEq1}
\sum_i \Pi_+ \, P_i \,\mathcal{F}_2 \, \rho^{\otimes 2} \, \mathcal{F}_2^\dagger \, P_i\, \Pi_+ + \sum_i \Pi_-\mathcal{F}_2 \, P_i \, \rho^{\otimes 2} \, P_i\,\mathcal{F}_2^\dagger \Pi_-\,.
\end{equation}
\item The probability that one measures $\mathcal{R}_2$ as $+1$, after having measured $\rho^{\otimes 2}$ to be in the subspace corresponding to $P_i$, is denoted by $\text{Prob}(+\, | \, i)$.  Similarly, the probability that one measures $\mathcal{R}_2$ as $-1$, after having measured $\rho^{\otimes 2}$ to be in the subspace corresponding to $P_i$, is denoted by $\text{Prob}(-\, | \, i)$.  After obtaining $\text{Prob}(+\, | \, i)$ and $\text{Prob}(-\, | \, i)$, one can compute
\begin{align}
\sum_i \lambda_i \bigg( \text{Prob}(+\, | \, i) - \text{Prob}(-\, | \, i)\bigg) &= \sum_i \lambda_i \, \text{tr}\left\{\Pi_+ \, P_i \,\mathcal{F}_2 \, \rho^{\otimes 2} \, \mathcal{F}_2^\dagger \, P_i\, \Pi_+ - \Pi_-\mathcal{F}_2 \, P_i \, \rho^{\otimes 2} \, P_i\,\mathcal{F}_2^\dagger \Pi_-\right\} \nonumber \\
&= \sum_i \lambda_i \, \text{tr}\left\{\mathcal{R}_2 \, P_i \,\mathcal{F}_2 \, \rho^{\otimes 2} \, \mathcal{F}_2^\dagger \, P_i\,\right\} \nonumber \\
&= \text{tr}\{\mathcal{R}_2 \, \mathcal{X}_2 \,\mathcal{F}_2 \, \rho^{\otimes 2} \, \mathcal{F}_2^\dagger \} \nonumber \\
&= \text{tr}\{X \, \rho^2\}\,,
\end{align}
where we have used $[\mathcal{X}_2,\mathcal{R}_2] = 0$ to go from the second line to the third line, and Eqn.~\eqref{R2appEq1} to go from the third line to the last line.  A similar procedure can be used to determine $\text{tr}\{\rho^2\}$, and then one can compute the quotient $\text{tr}\{X\, \rho^2\}/\text{tr}\{\rho^2\}$.
\end{enumerate}

In an actual experiment, one does not directly measure the parity operator $\mathcal{R}_2$, but instead measures the number operator on every site.  Since the common refinement of the eigenspaces of all of the number operators is a refinement of the eigenspaces of $\mathcal{R}_2$, one can measure $\mathcal{R}_2$ via the number operators and obtain the same result as above.

\subsection{Fermion interferometry}

It is straightforward to adapt the boson interferometry techniques to fermions, although a few modifications to the protocol are required.  Our protocol is inspired by the work of \cite{Pichler1}. Suppose we have two systems of fermions, and require that states of different fermion number lie in different superselection sectors.  Technically, the superselection rule means that for all observables $X$, we have $\langle \psi_1 | X |\psi_2\rangle = 0$ if $|\psi_1\rangle$ and $|\psi_2\rangle$ are states of definite, but distinct fermion number.

For fermions, it is \textit{not} true that $\text{tr}\left\{\mathcal{R}_2 \, \mathcal{F}_2 \, \rho^{\otimes 2} \, \mathcal{F}_2^\dagger \right\} = \text{tr}\{\rho^2\}$.  Instead, we have
\begin{equation}
\label{fermionEq1}
\text{tr}\left\{\mathcal{V} \, \mathcal{F}_2 \, \rho^{\otimes 2} \, \mathcal{F}_2^\dagger \right\} = \text{tr}\{\rho^2\}
\end{equation}
where $\mathcal{V}$ has eigenvalues $\pm 1$ which depend on the total number of fermions $N_{\text{tot}}$, the floor of half of the total number of fermions $\lfloor N_{\text{tot}}/2 \rfloor$, and the number of fermions $N_2$ in the second copy of the subsystem.  (There are, in fact, many choices of $\mathcal{V}$ which satisfy Eqn.~\eqref{fermionEq1}, and so we choose a convenient one for our purposes.)  The measurement outcomes for $\mathcal{V}$ are given in the table below: \\
\begin{table}[h]
\begin{tabular}{ c | c | c | c }
  $N_{\text{tot}}$ & $\lfloor N_{\text{tot}}/2 \rfloor$ & $N_2$ & Result \\  
  \hline			
  Even & Even & Even & $+1$ \\
  Even & Even & Odd & $-1$ \\
  Even & Odd & Even & $-1$ \\
  Even & Odd & Odd & $+1$ \\
  Odd & Even & Even & $+1$ \\
  Odd & Even & Odd & $-1$ \\
  Odd & Odd & Even & $-1$ \\
  Odd & Odd & Odd & $+1$ \\
\end{tabular}
\caption{Characterization of measurement outcomes for $\mathcal{V}$.}
\end{table}
\vskip.3cm
The procedure for measuring $\text{tr}\{\mathcal{X}_2\, \rho^2\}/\text{tr}\{\rho^2\}$ is the same as in the bosonic case above, except that now we need 
\begin{equation*}
\mathcal{X}_2 = \mathcal{F}_2\left(\frac{1}{2}\,X(\{f_{i,1}, f_{i,1}^\dagger\}) + \frac{1}{2}\,X(\{f_{i,2}, f_{i,2}^\dagger\})\right)\mathcal{F}_2^\dagger\,,
\end{equation*}
(where here the $f,f^\dagger$ operators are fermionic) to additionally satisfy
\begin{equation}
[\mathcal{X}_2, \mathcal{V}] = 0\,.
\end{equation}
So first let us find which operators, in general, commute with $\mathcal{V}$.  Suppose we have an operator of the form
\begin{equation}
\underbrace{f_{i_1,1}^\dagger \cdots f_{i_{m_1},1}^\dagger}_{m_1\text{ of these}} \, \underbrace{f_{j_1,1} \cdots f_{j_{m_2},1}}_{m_2\text{ of these}} \, \underbrace{f_{k_1,2}^\dagger \cdots f_{k_{n_1},2}^\dagger}_{n_1\text{ of these}} \, \underbrace{f_{\ell_1,2} \cdots f_{\ell_{n_2},2}}_{n_2\text{ of these}}
\end{equation}
where $\{i_1,...,i_{m_1}\}$, $\{j_1,...,j_{m_2}\}$, $\{k_1,...,k_{n_1}\}$, $\{\ell_1,...,\ell_{n_2}\}$ are all sets with \textit{non-repeating} elements.  All of these operators transform multiplicatively by either $+1$ or $-1$ after conjugation by $\mathcal{V}$.  Letting $m = |m_1-m_2|$ and $n = |n_1 - n_2|$, the possibilities are tabulated below: 
\\
\begin{table}[h]
\begin{tabular}{ c | c | c | c }
  $m+n \,\,\, (\text{mod\,}2)$ & $m+n \,\,\, (\text{mod\,}4)$& $n\,\,\, (\text{mod\,}2)$ & Result \\  
  \hline			
  $0$ & $2$ & $0$ & $-1$ \\
  $0$ & $2$ & $1$ & $+1$ \\
  $0$ & $0$ & $0$ & $+1$ \\
  $0$ & $0$ & $1$ & $-1$ \\
  $1$ & $1$ & $0$ & $-1$ \\
  $1$ & $1$ & $1$ & $+1$ \\
  $1$ & $3$ & $0$ & $+1$ \\
  $1$ & $3$ & $1$ & $-1$ \\
\end{tabular}
\caption{Transformation of products of fermion operators under conjugation by $\mathcal{V}$.}
\end{table}
\vskip.5cm
For example, letting $\mathcal{X}_2 = \frac{1}{2}(n_{i,1} + n_{i,2})$, we have $[\mathcal{X}_2, \mathcal{V}] = 0$.  If instead $\mathcal{X}_2 = \mathcal{F}_2(\frac{1}{2}(n_{i,1} n_{j,1} + n_{i,2} n_{j,2}) \mathcal{F}_2^\dagger$, we likewise have $[\mathcal{X}_2, \mathcal{V}] = 0$.

\section*{II. \quad Extracting two point correlations in the low temperature limit}
In this section, we numerically study the effect of the second term in Eqn.~\eqref{unconventional1} in the main text.
More specifically, we have argued that one can extract a density-density correlation from a more experimentally accessible quantity:
\begin{align}
C(j,\ell) \equiv 
\frac{1}{2} \textrm{tr} \left\{ n_j \, n_\ell \,\rho(T/2) \right\} 
+ \frac{1}{2} \frac{\textrm{tr}\left\{ n_j \,\rho(T)\, n_\ell \,\rho(T) \right\}}
{\textrm{tr}\left\{ \rho(T)^2 \right\}}.
\end{align}
While the first term is the desired density-density correlation, the second term arises as a consequence of the Fourier transform of local operators $n_j$ and $n_\ell$. As described in the main text, however, we expect that at sufficiently low temperatures the second term does not exhibit any systematic dependence on the distance between two points $d = |j -\ell|$, allowing us to extract physically meaningful quantities such as correlations lengths from fitting $C(j,\ell)$ as a function of $d$. 

In order to confirm this expectation, we consider a 1D Bose-Hubbard Hamiltonian with nearest-neighbor hopping rate $J$ and on-site repulsive interaction $U=3J$. We numerically compute  thermal density matrices for $N=4$ particles on $L=16$ lattice sites with periodic boundary condition at various temperature $T/J \in \{\frac{1}{10}, \frac{1}{5}, \frac{1}{4}, \frac{1}{2}, 1\}$. For each temperature $T$, we compute each term in $C(j,\ell)$ as well as their sum as a function of the distance $d\in \{1, \dots, 8\}$.
\begin{figure}[h!]
  \centering
  \includegraphics[width=6in]{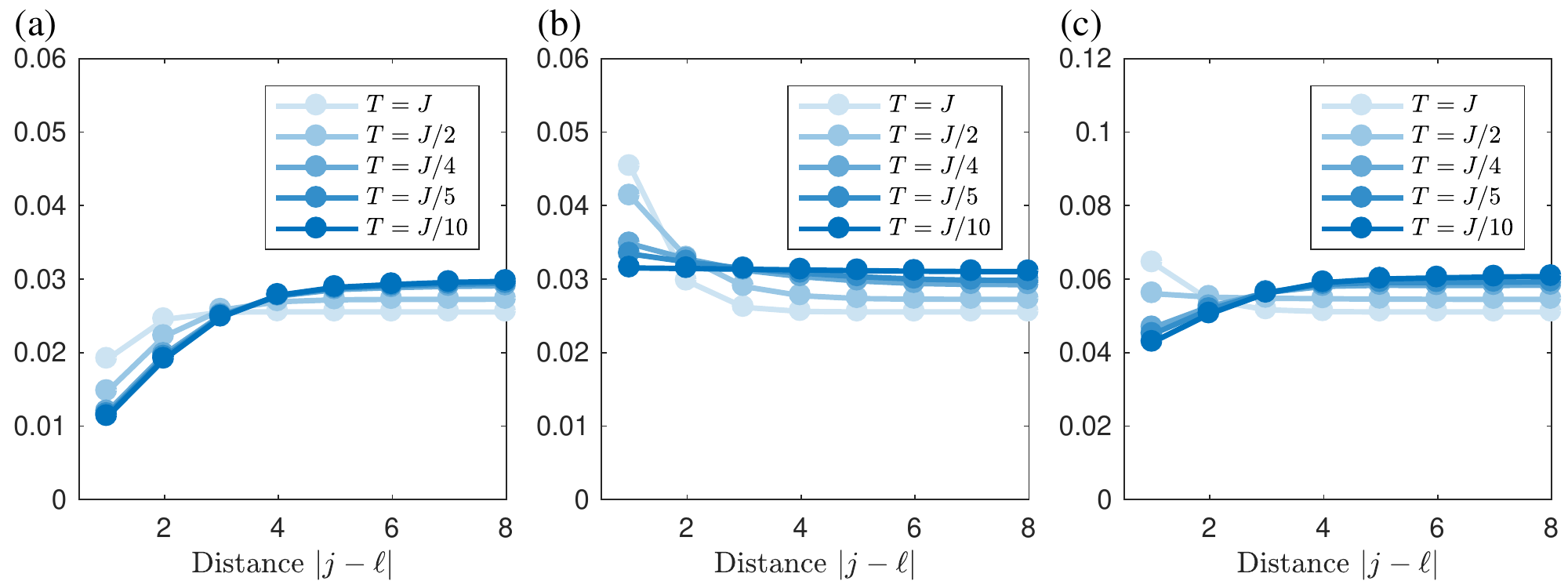}
  \caption{Extracting the density-density correlation from Eqn.~\eqref{unconventional1} in the main text. (a) Density-density correlations in a 1D Bose-Hubbard model at various temperatures. This quantity corresponds to the first term in $C(j,\ell)$. (b) Additional contribution to $C(j,\ell)$ arising from the second term. Crucially, this contribution exhibits decreasing distance-dependence in the low temperature limit. (c) The position dependence of the total value $C(j,\ell)$ is dominated by the first term in low temperature limit. }
  \label{fig:spurious_term}
\end{figure}
Fig.~\ref{fig:spurious_term} below
%HERE
summarizes our numerical results, from which it can be checked that the density-density correlation (the first term in $C(j,\ell)$) displays strong anti-bunching (Fig.~\ref{fig:spurious_term}a) at low temperature. By contrast, the second term exhibits diminishing distance-dependence as the temperature decreases (Fig.~\ref{fig:spurious_term}b). We find that the distance dependence of the total value $C(j,\ell)$ is indeed dominated by the density-density correlation (Fig.~\ref{fig:spurious_term}c) at sufficiently low temperatures.

\section*{III. \quad Experimental methods}
Our experiments start from a high fidelity Mott insulator with a single particle per lattice site. Using high-precision, site-resolved optical potentials, created by a digital micro-mirror device (DMD), we isolate two neighboring six-site long chains of atoms with exactly one atom on each site. In order to ensure the high fidelity of the initial state we hold it in the $45 \mathrm{E_r}$ deep optical lattice in both directions. To obtain a locally thermal state we suddenly drop the lattice depth along the chains, allowing atoms to tunnel, while keeping the lattice high between the chains. We use a pair of DMD beams to offset the sites right outside the region of interest, thereby defining the overall length of the system. After variable evolution time, we freeze the dynamics along the chains by suddenly ramping up the lattice back to $45 \mathrm{E_r}$.  In order to make sure that the state has thermalized, we pick evolution times for which the entanglement entropy of the region of interest has reached its saturation value. Table~\ref{tab:times} shows the times used in Fig.~\ref{fig2} in the main text for each case studied. 

\begin{table}[h]
	
	\begin{tabular}{|c|c|}
		\hline Case & Times ($\hbar/J$)\\
		\hline A & 1.0, 1.4, 2.2, 4.3, 5.1, 6.4, 8.4 \\ 
		\hline B & 12.2, 24.0, 59.4 \\ 
		\hline C & 22.4, 41.3 \\ 
		\hline
	\end{tabular} 
	\caption{Evolution times used for each case in Fig.~\ref{fig2} of main text.}
	\label{tab:times}
\end{table}

In order to implement the beamsplitter operation, we drop the lattice depth between the chains and let the atoms evolve for a certain time duration. During this process the lattice depth along the chains stays high, preventing wavefunction evolution in that direction. At the end of this sequence, we read out the state of the system in the particle number basis with single-site and full atom-number resolution. For more details see \cite{Kaufman2016}. 

% figure from Thermalization supplementary maybe?
% \subsection{Data analysis}

\end{document}